\documentstyle[12pt,epsf]{article}

\begin{document}
\begin{titlepage}
\title{\vskip -1.5cm
\hfill
\parbox{4cm}{\normalsize DAMTP 96-45}\\
\vspace{1cm}
$\phi^4$ Kinks - Gradient Flow and Dynamics}
\author{N. S. Manton\thanks{e-mail address:
\tt N.S.Manton@damtp.cam.ac.uk}
 ~and~  H. Merabet\thanks{e-mail address: \tt M.Houari@damtp.cam.ac.uk}
\\
{\normalsize\it
Department of Applied Mathematics and Theoretical Physics,
University of Cambridge}\\
{\normalsize\it
Cambridge CB3 9EW, United Kingdom}}
\date{\normalsize May 1996}
\maketitle
\thispagestyle{empty}

\vspace*{-1cm}
\begin{abstract}
\normalsize
The symmetric dynamics of two kinks and one antikink in classical
(1+1)-dimensional $\phi^4$ theory is investigated.  Gradient flow
is used to construct a collective coordinate model of the system.
The relationship between the discrete vibrational mode of a single
kink, and the process of kink-antikink pair production is explored.

\end{abstract}

\end{titlepage}

\section{Introduction}

One of the more challenging problems in non-integrable soliton
dynamics is to find a collective coordinate model
of the dynamics. In some theories there is a moduli space of exact
multi-soliton static solutions, and the soliton dynamics at low energy
is modelled by geodesic motion on the moduli space 
\cite{Manton82,Stuart95}. In others there is a nearby theory
with exact static solutions and a moduli space, and the soliton
dynamics is modelled by geodesic motion on the moduli space perturbed
by small forces \cite{Shah94,Stuart94}.
In exactly integrable models, like the sine-Gordon model, one can
exactly compute what happens in a multi-soliton or
multi-soliton-antisoliton process.  There is no energy
loss to radiation in these processes, and solitons do not
annihilate antisolitons in a collision.
However, the problem of interest here involves soliton-antisoliton
annihilation. There is no moduli space of static solutions
available; neither is it close to being exactly integrable.

A method of dealing with solitons, or solitons and antisolitons, 
which exert forces on each
other, is constrained minimization \cite{Jac/Reb79}. One introduces
collective coordinates by hand, say the positions of the solitons and
antisolitons, and  then seeks the field configuration of minimal
potential energy for given positions. The reduced  model for
the soliton dynamics is partly defined by this (minimal) potential
energy as a function of the collective coordinates.
There are various disadvantages of this procedure. Firstly, it is
rather {\it ad hoc}, because there are many possible definitions of
the collective coordinates. When solitons are close together, their
positions are not uniquely defined. Different definitions will lead
to different reduced models. Secondly, the kinetic energy is not
well-defined. It is tempting to use the naive kinetic energy
expression, based on the soliton masses and the collective coordinate
velocities, but this is again quite misleading when the solitons are
close together. We have learnt from examples like BPS monopoles, where
there is a moduli space of exact static solutions, that the kinetic
energy is determined by the geometry of the moduli space, and it
could not easily be determined if one had a rigid {\it a priori} view of
what the collective coordinates of the monopoles were.

The method we use here, advocated in \cite{Manton88} and previously
developed in the context of
instanton-antiinstanton physics in \cite{Bal/Yun86}, exploits the
gradient flow equations associated with the field theory of interest.
The detailed procedure is explained below. Here let us point out that
gradient flow represents a dissipative dynamics of fields, and hence
of solitons and antisolitons. But this gradient flow is only used as
an auxillary tool to construct a non-dissipative reduced dynamical
system, with its own Lagrangian.
The gradient flow curves are chosen to map out a finite dimensional
manifold of field configurations, and the reduced Lagrangian is simply
the field theory Lagrangian restricted to this manifold. A natural
choice for the gradient flow curves is the union of all those that
descend from a particular unstable, static solution of the field
equations. This defines the unstable manifold of the solution.
Alternatively, for solitons that attract each other, one
can define the unstable manifold as the union of all the gradient flow
curves which descend from configurations of very well separated solitons.

The unstable manifold acquires a metric and potential from the field
configuration space of which it is a submanifold. These
define a Lagrangian for dynamical motion on the
unstable manifold, leading to an Euler-Lagrange equation of the
form (\ref{natEL}) below. This reduced dynamical system
is the collective coordinate model for the soliton dynamics. Any set
of coordinates on the unstable manifold can be regarded as collective
coordinates for the solitons, although there may be a preferred choice
if the solitons are well-separated. A
solution of the Euler-Lagrange equation on the unstable manifold
is not, in general, identical with any solution of the original field
equations, but it may provide an approximation.

The reduced system should be particularly useful if the potential
energy function does not vary much over the unstable manifold,
that is, if the static forces between
solitons are weak (but nonvanishing), so large kinetic energies are
not generated. The unstable manifold is then an almost flat valley in
the field configuration landscape, and the motion on it at modest speeds is
expected to be a good approximation to the true field dynamics. This
is because motion orthogonal to the valley, i.e. radiation, is not
readily generated. However, this has not really been
demonstrated analytically, nor tested numerically. In the example to
be considered here, the potential energy varies substantially, but we
shall show that the reduced dynamics is still useful.

The gradient flow in field theories with solitons has been investigated in
detail in some cases. Demoulini and Stuart have studied the
gradient flow in
the two-dimensional abelian Higgs model \cite{Dem/Stu96}, whose
solitons are magnetic flux vortices, proving the global
existence of gradient flow curves, and obtaining some insight into the
unstable manifolds. For $N$ type II vortices, which repel, the unstable
manifold of the static axisymmetric solution representing $N$
coincident vortices is particularly interesting. The reduced
dynamics on this manifold should be a good model for $N$-vortex
motion, but this
has not been investigated. Waindzoch and Wambach have studied the
gradient flow in the two-Skyrmion sector of Skyrme's soliton model of
nucleons \cite{Wai/Wam95},
the geometry of which was discussed in \cite{Ati/Man93}, and have
numerically (partially) constructed an unstable manifold which should model
the two-Skyrmion dynamics. Also, there has been a detailed study of the
gradient flow equation in $\phi^4$ theory with $\phi$
defined on a finite spatial interval \cite{H/M/O/83}.
It has been shown that the unstable manifold of the simplest unstable
solution on the interval, namely $\phi =0$, completed with the
unstable manifolds of lower energy solutions, is a global attractor for
the gradient flow.
However, there has been no comparison made between the solutions
of the $\phi^4$ theory field equation and the dynamical motion
on the unstable manifold.
This example of $\phi^4$ theory on an interval shows that an unstable
manifold can have a non-smooth boundary
(see Fig. 8.5 in ref. \cite{H/M/O/83}).
For an unstable manifold to be useful for modelling multi-soliton
dynamics, and especially quantized soliton dynamics, it is important
that the manifold has no boundary at all. The vortex and Skyrmion
examples appear to satisfy this requirement. Mathematically, it is not
at all clear in what circumstances an infinite-dimensional gradient
flow system has a smooth, finite-dimensional global
attractor without boundary.

Here we apply the gradient flow method to a particular case of
kink-antikink dynamics in $\phi^4$ theory defined on an infinite line.
The reduced system is one-dimensional
and describes a process by which the kink and antikink annihilate,
exciting a particular mode of the field. In the reduced system, this
process then reverses and the kink-antikink pair is recreated.
Since we are interested in how the kink and antikink annihilate
into radiation, we perform a further analysis of how the reduced
system couples to the radiation modes, and estimate the rate at which the
energy of the reduced system is transfered to radiation. Fortunately,
this rate is relatively slow, and we find a good agreement
between the complete field dynamics and the dynamics of the reduced
system coupled weakly to radiation.

\section{$\phi^4$ Theory}
The model we consider is $\phi^4$ theory on a line.
This model is a nonlinear, Lorentz invariant, scalar field theory
which is not integrable \cite{Rajaraman82}. The field $\phi(x,t)$ has
kinetic energy
\begin{equation}\label{kinetic}
T = \int_{-\infty}^{\infty} {1 \over 2} \dot\phi^2  \ dx
\end{equation}
and potential energy
\begin{equation}\label{potential}
V = \int_{-\infty}^{\infty} ({1\over 2} \phi'^2 + {1\over 2} (\phi^2 -
1)^2) \ dx \ .
\end{equation}
The Lagrangian is $L=T-V$, and this leads to the field equation
\begin{equation}\label{full-eq}
\ddot\phi - \phi '' + 2\phi \ (\phi^2 -1) = 0.
\end{equation}
There are two vacua, $\phi = \pm 1$.  There are also two types of
solitons, called kinks and antikinks. A kink interpolates
between $-1$ and $1$ as $x$ increases, and an antikink does the
reverse.
Eq.(\ref{full-eq}) has static solutions
\begin{equation}\label{kink}
\phi_K (x-a) = \tanh (x-a)
\hspace{0.3cm}  \mbox{and} \hspace{0.3cm}
\phi_{\bar K} (x-b) = -\tanh (x-b),
\end{equation}
where $\phi_K$ and $\phi_{\bar K}$ are the kink
and the antikink respectively.
They have potential energy (rest mass) ${4\over 3}$.
The parameter $a$ ($b$) denotes the centre of the kink (antikink).
These solutions can be Lorentz boosted to an arbitrary speed $v < c$
(where the speed of light $c=1$ in our units). There are non-static
field configurations with several kinks and antikinks, but kinks and
antikinks must alternate along the line. A neighbouring kink and
antikink can annihilate into oscillations of the field
$\phi$ -- interpreted as radiation. Also, in suitable circumstances, a
kink-antikink pair can be produced.

There has been a substantial investigation of kink-antikink dynamics
(see, e.g. ref.\cite{C/S/W/83}).
If a kink and antikink approach from infinity, starting with any
speed much less than the speed of light, they will annihilate into
radiation.  In higher speed collisions, what happens depends
sensitively on the speed.

We have investigated a different process, which appears to be better
suited to a collective coordinate analysis.  This is the symmetric
motion of a configuration of two kinks and an antikink, which we
denote $K\bar{K}K$.
We consider fields which possess the reflection symmetry
\begin{equation}\label{symmetry}
\phi (x) = - \phi (-x)
\end{equation}
at all times, and for which $\phi \rightarrow 1$ as $x \rightarrow \infty$.
This allows, among other things, for there to be kinks
centred at $\pm a$, and an antikink at the origin.  A
field of this type can evolve into a single kink plus radiation, but
the radiation is not emitted very rapidly, as we shall show.
In the case of pure $K\bar{K}$ annihilation, all the energy goes
into radiation \footnote{This is also a rather slow process because
an intermediate, long-lived breather state is produced \cite{C/S/W/83} .}

An interesting and important feature of $\phi^4$ kink dynamics is that
a single kink has a normalizable discrete vibrational mode.
The discrete mode deforms a static kink at the origin to the form
\begin{equation}\label{breather}
\phi (x) = \phi_K (x) + A \ \eta_D (x)
\hspace{0.3cm} \mbox{with} \hspace{0.3cm}
\eta_D (x) = {\sinh (x) \over \cosh^2(x)} \ ,
\end{equation}
where, in the linearized approximation, the amplitude $A$ oscillates
with frequency $\omega_0 = \sqrt{3}$. In comparison,
the continuum radiation modes have frequencies $\omega > 2$.
Fields of the form (\ref{breather}) have the
reflection symmetry (\ref{symmetry}).  When $A \simeq -2$, the
field (\ref{breather}) looks rather like a
$K\bar{K}K$ configuration with the kinks close together.
In fact it is possible to smoothly interpolate between a $K\bar{K}K$
configuration with the kinks well-separated, and a single kink,
via fields of the approximate form (\ref{breather}).

\section{Gradient Flow}
\setcounter{equation}{0}

We recall that a natural (finite-dimensional) Lagrangian system on a
configuration space $\cal{C}$ has
a Lagrangian of the form \cite{Arnold78}
\begin{equation}\label{natLag}
L({\dot {\bf{y}}},{\bf{y}}) =
T({\dot {\bf{y}}},{\bf{y}}) -
V({\bf{y}}) \ ,
\hspace{0.5cm} \mbox{where} \hspace{0.5cm}
T({\dot {\bf{y}}},{\bf{y}}) =
{1\over 2} g_{ij}({\bf{y}})
\dot{y}^i \dot{y}^j \ .
\end{equation}
Here, $y^i$ are arbitrary coordinates on $\cal{C}$,
and $g_{ij}({\bf{y}})$ can be interpreted as a Riemannian
metric on $\cal{C}$. The Euler-Lagrange equation is
\begin{equation} \label{natEL}
\frac{\mbox{d}}{\mbox{d}t} (g_{ij} \dot{y}^j)
- {1\over 2} \frac{\partial g_{jk}}{\partial y^i}
\dot{y}^j \dot{y}^k = - \frac{\partial V}{\partial y^i}
\end{equation}
whereas the gradient flow equation is
\begin{equation} \label{natgradflow}
g_{ij} \dot{y}^j =
- \frac{\partial V}{\partial y^i} \ .
\end{equation}
The gradient flow is orthogonal to the contours of the potential $V$
-- the notion
of orthogonality requires a metric -- and in the direction of
decreasing $V$. If ${\bf y}_0$ is a saddle point of $V$, then the
unstable manifold of ${\bf y}_0$ is defined as the union of all the
gradient flow curves that descend from ${\bf y}_0$.
If the configuration space is Euclidean, and
$y^i$ are Cartesian coordinates, then the gradient flow equation
is obtained from the Euler-Lagrange equation by replacing second
time derivatives by first time derivatives. 

$\phi^4$ field theory
can be thought of as an infinite-dimensional Lagrangian system whose
configuration space $\cal{C}$ is the space of fields $\{ \phi(x) \} $
at a given time. The simple form of the kinetic energy expression
(\ref{kinetic}) implies that $\cal{C}$ is Euclidean, with the
Riemannian distance between fields $\phi(x)$ and $\phi(x) +
\delta\phi(x)$ being the square root of
\begin{equation} \label{Riem}
\int_{-\infty}^{\infty} (\delta\phi(x))^2 \ dx \ .
\end{equation}
The gradient flow
equation is therefore obtained from eq.(\ref{full-eq}) simply by replacing
$\ddot{\phi}$ by $\dot{\phi}$, giving
\begin{equation}\label{flow-eq}
\dot{\phi} = \phi'' - 2\phi \ (\phi^2 -1).
\end{equation}
Here, we solve this gradient flow equation to construct a collective
coordinate model for the $K\bar{K}K$ system, with the imposed 
reflection symmetry (\ref{symmetry}).
Because of this symmetry, the $K\bar{K}K$ system is modelled
by a one-dimensional unstable manifold. It consists of the field configurations
$\phi(x)$, parametrized by $-\infty < t < \infty$, occurring
in the solution of the gradient flow equation (\ref{flow-eq}) which
descends from a $K\bar{K}K$ configuration,
with kinks at $\pm \infty$ and an antikink at the origin, to a single
kink at the origin. This solution approaches the single kink tangent to its
discrete mode. The collective coordinate can be identified either with
the separation of the kinks from the antikink, while this is large,
or with the discrete mode amplitude, while this has
a modest or small negative value.

The unstable manifold is infinitely long.  At the top end,
where there are kinks at infinite
separation, the potential energy is $4$ (three times the kink rest
mass).  At the bottom, where there is a single kink, the
potential energy is ${4 \over 3}$.
Most of the unstable manifold consists of well-separated
$K\bar{K}K$ configurations, with kinks at $\pm a$ and an antikink at
the origin.  We cannot write down the exact form of the
fields satisfying (\ref{flow-eq}), but for $a >> 1$ an exponentially accurate
approximation is given by the product form
\begin{equation}\label{initial-field}
\phi  = - \phi_K (x + a) \ \phi_{\bar K} (x) \ \phi_K (x - a) \ ,
\end{equation}
with $a$ a function of time.  To a sufficiently good approximation,
the Lagrangian
for such fields is ${4\over 3} \dot a^2  + 32 e^{-2a} - 4$, since
there are two kinks moving with velocities $\pm {\dot a}$, and the interaction
potential of a $K\bar{K}$ pair separated by $a$ is $-16 e^{-2a}$ if $a$
is large \cite{Manton79}. The gradient flow equation therefore reduces to
\begin{equation}\label{a-flow}
\dot{a} = -24 e^{-2a}
\end{equation}
whose integral is $e^{2a} = -48t  + \hbox {const}$.  We have verified
this by taking the initial data (\ref{initial-field}) with $a=5$,
and evolving it numerically with the gradient flow equation
(\ref{flow-eq}) until the three zeros of $\phi$ are at $x = \pm 4$
and $x=0$.  The time this takes is very close to
${1\over 48} (e^{10} - e^{8}) \simeq 397$,
which is a very long time compared with the period
of the discrete mode oscillation.

For $a < 4$, the expression (\ref{initial-field}) fails to be a good
approximation for the fields on the unstable manifold, so we have
integrated (\ref{flow-eq}) numerically, starting at $t=0$
with initial data (\ref{initial-field}) with $a=4$. We find that after
a finite time the $K\bar{K}K$ configuration annihilates into a
deformed single kink. An exact annihilation
time can be defined as the time when the three zeros of $\phi$
coalesce into one. Equivalently, this is when $\phi'$ is zero at
the origin. ($\phi'$ is negative before this time and positive
afterwards.) The annihilation time, for our initial data, is $t=59.64$.
By integrating (\ref{a-flow}) and finding
when $a=0$, one may estimate the annihilation time to be
$t = {1\over 48} (e^8 -1) \simeq 62$.

As $t$ increases further, the field configuration rapidly approaches the
single kink, and it does so tangent to the discrete mode, since this
is the lowest frequency mode of vibration around the kink. So the
field becomes well approximated by the form (\ref{breather}), with an
amplitude $A(t)$ tending to zero. Even at
the annihilation time, the field (\ref{breather}) with $A=-1$ is a
reasonable approximation, since for this value of $A$, $\phi'$
vanishes at the origin.
The asymptotic behaviour of $A(t)$ is given by calculating from
(\ref{kinetic}) and (\ref{potential}) the
kinetic and potential energies for fields of the form (\ref{breather}).
These are, respectively,
\begin{equation}\label{A-Kin}
T (A) = {1\over 3} \dot{A}^2
\end{equation}
and
\begin{equation}\label{A-Pot}
V (A) = {4\over 3} + A^2 + {\pi\over 8} A^3 + {2\over 35} A^4,
\end{equation}
giving a gradient flow equation
\begin{equation}\label{A-flow}
{2\over 3}\dot{A} = -2 A - {3\pi\over 8} A^2 - {8\over 35} A^3.
\end{equation}
As $t \rightarrow \infty$, $A$ decays like $e^{-3t}$, the coefficient being the
square of the discrete mode frequency.

\begin{figure}[htc]
\centerline{ \epsfxsize=10cm \epsfbox{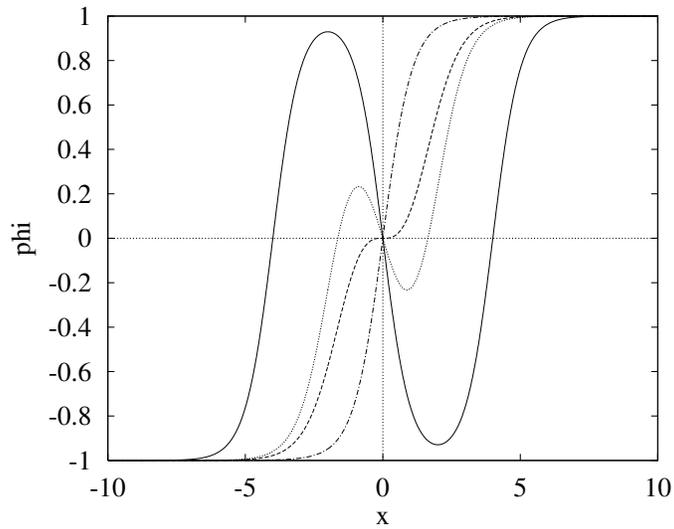} }
\caption{The gradient flow annihilation process for the $K\bar{K}K$
system at time 0 (solid), 59 (dotted), 59.64 (dashed) and 61 (dashed dotted).}
\label{fieldsflow:fig}
\end{figure}

\begin{figure}[htc]
\centerline{ \epsfxsize=10cm \epsfbox{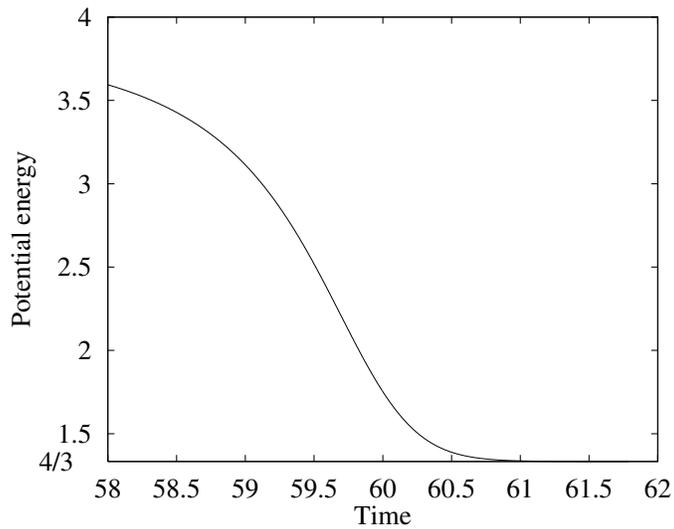} }
\caption{ Time dependence of the potential energy along the gradient
flow curve. }
\label{potflow:fig}
\end{figure}

Our numerical solution of the gradient flow equation
was obtained using a predictor-corrector
finite difference scheme. Since this is an implicit scheme,
there is no stability restriction on $\Delta t/\Delta x^2$
(see \cite{Ames92}). The resulting tridiagonal linear system is solved
explicitly using the Thomas algorithm. Our space-time domain
is defined on  $- 10 \leq x \leq 10$ and $0 \leq t \leq 65$
in space and time steps of 0.02 and 0.005 respectively.
 Fig. \ref{fieldsflow:fig} shows the field $\phi$ at the initial time
and various subsequent times. The annihilation process occurs at $t\sim 60$.
Fig. \ref{potflow:fig} shows the potential energy as a function of time.
The rapid decrease occurs close to the annihilation time.

The unstable manifold stops abruptly at the single kink.
It is desirable to define a valley which
smoothly continues it. The valley continuation needs to approach the
single kink tangent to the discrete mode from the opposite direction.
The best continuation we can think of is the set of fields
(\ref{breather}) with $A > 0$.  With this ansatz,
the potential energy in the valley, and its first, second and third
derivatives, are all continuous at the single kink, but the fourth derivative
is probably discontinuous. The positive value
of $A$ for which the potential energy expression (\ref{A-Pot})
has value 4 is 1.3.

Any field $\phi (x)$, with the
reflection symmetry and boundary conditions we are assuming, has a
decomposition
\begin{equation}\label{kink-field}
\phi (x) = \phi_K (x) + A \ \eta_D (x) + \eta (x)
\end{equation}
where $\eta (x)$ is a superposition of continuum modes orthogonal
to the discrete mode $\eta_D (x)$.  We can calculate the amplitude of
the discrete mode $A$ by projection
\begin{equation}\label{amplitude}
A  = {3 \over 2} \int^{\infty}_{-\infty}
\left ( \phi (x) - \phi_K (x) \right ) \eta_D (x) \mbox{d}x \ .
\end{equation}
In the valley, $A$ varies between $-3 \pi/2$ and $\infty$.
The first number is calculated using (\ref{amplitude}) with
$\phi(x)$ a $K\bar{K}K$ configuration with kinks at
infinite separation, and its finite value implies that the tangent
to the unstable manifold is asymptotically orthogonal to the discrete
mode (both metrically, and in the usual sense of integration).
The $K\bar{K}K$ half of the valley is therefore curved, turning a right angle
along its length. The half of the valley defined using the
ansatz (\ref{breather}), with $A$ positive, is straight.

\begin{figure}[htc]
\centerline{ \epsfxsize=10cm \epsfbox{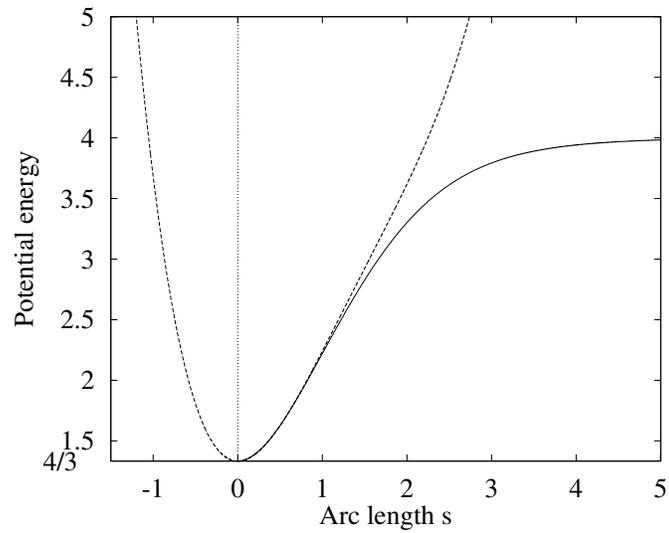} }
\caption{ Arc length dependence of the potential energy in the valley.
Gradient flow (solid) and discrete mode results for $s<0$
(the discrete mode result for $s>0$ is also shown).}
\label{spotflow:fig}
\end{figure}

Along the valley, the arc length $s$ is the only intrinsic geometrical
quantity. Suppose $s = 0$ at the
bottom, with $s$ positive in the $K\bar{K}K$ half of the valley.
Let $V(s)$ be the potential energy in
the valley, as a function of arc length. The
intrinsic gradient flow equation is
\begin{equation}\label{s-flow}
\dot{s} = - {\mbox{d}V \over \mbox{d}s}
\end{equation}
and in the $K\bar{K}K$ half of the valley this must give the same
dependence of $V$ with time,
as the field gradient flow equation. Eq. (\ref{s-flow}) implies
\begin{equation}
{\mbox{d}V\over \mbox{d}t} =
 - \left ( {\mbox{d}s\over \mbox{d}t} \right )^2 \ .
\end{equation}
We have taken our numerically obtained $V(t)$ and integrated
$\sqrt{-\dot{V}}$ to find $s(t)$, and hence determined $s$ and $V(s)$
along the curved half of the valley. The kinetic energy expression
(\ref{A-Kin}) for a field of the form (\ref{breather}) implies that
for small $s$, we may identify $s = -\sqrt{2\over 3}A$. For large
positive $s$ we may identify $ds = \sqrt{8\over 3} da$.  Numerically we
estimate that for large $a$, $s= \sqrt{8\over 3} a - 1.18$.
The arc length along the straight half of the valley is exactly
$s = -\sqrt{2\over 3}A$, and using this relation we can easily
convert the potential function (\ref{A-Pot}) into a function of $s$.
Fig. \ref{spotflow:fig} shows the potential function $V(s)$ in both
halves of the valley.
If we use the discrete mode ansatz (\ref{breather}) also for $s > 0$ we
get the dashed curve in the Figure, which is not a bad approximation
for $|A| \leq 2$.
Note the asymmetry of the potential, due to the $A^3$ term in $V$.

\section{Dynamics}
\setcounter{equation}{0}

The one-dimensional dynamical system
\begin{equation}\label{s-lagrangian}
L = {1\over 2} \dot{s}^2 - V(s) \ ,
\end{equation}
with $V(s)$ as shown in Fig. \ref{spotflow:fig}, is our collective coordinate
model for the dynamics of the $K\bar{K}K$ system.  It predicts
that for energies ${4\over 3} < E < 4$, there is a nonlinear
oscillatory behaviour.  For $E > 4$ it predicts that two kinks
can approach from infinity and annihilate an antikink at the origin, with
the kinetic energy being captured by the discrete mode of the kink that
remains; then the process reverses and
$K\bar{K}K$ reform.  The model also predicts that if the initial data
is of the form (\ref{breather}) with $A=0$ and $|\dot{A}| > 2 \sqrt{2}$,
then a $K\bar{K}K$ configuration will form, and the kinks move off to
infinity.  (The more direct process is with $\dot{A}$ initially negative.)

We do not expect the field dynamics to be exactly along the part of
the valley defined by the gradient flow, since, as we have shown,
the valley is extrinsically curved. However, we expect the motion
along the valley to be a guide, rather like the bottom of a bobsleigh
run is a guide to the bobsleigh trajectories.  As the valley turns, so
the dynamical motion must climb up the side of the valley to be forced
round the corner.  We also expect some vibrational motion orthogonal
to the valley to be generated.

We have numerically studied the field dynamics of the $K\bar{K}K$
system, given by eq.(\ref{full-eq}).
The integration is performed using a second-order three
level implicit formula. As for the gradient flow case,
this scheme is stable for any value of $\Delta t/\Delta x$
provided that the three level parameter is greater than 1/4
(see \cite{Ames92}). Our space-time domain is defined on
$- 20 \leq x \leq 20$ and $0 \leq t \leq 60$
in space and time steps of 0.02 and 0.015 respectively.
We have considered two kinds of initial data.

\begin{figure}[htc]
\setlength{\unitlength}{1in}
\begin{picture}(4.6,2.3)(0,0)
\put(0.9,-0.2){\epsfbox{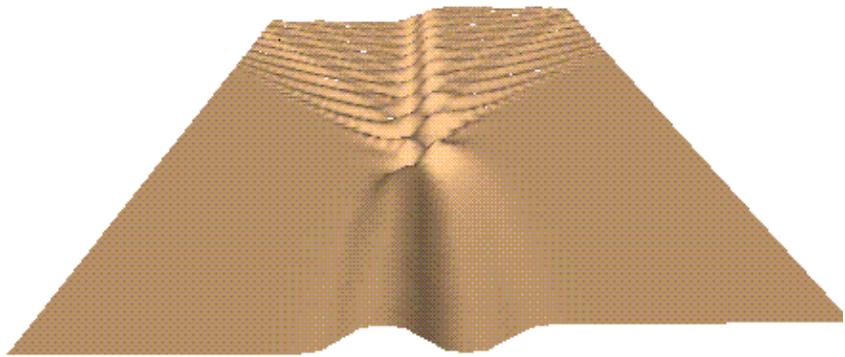}}
\end{picture}
\caption{ The full dynamics annihilation process. }
\label{annihfull:fig}
\end{figure}

First, we have taken the
initial data (\ref{initial-field}) with $a=4$ and with $\dot{\phi}=0$.
The initial potential energy is very close to 4.
We observe (see Fig. \ref{annihfull:fig})
that the two kinks approach the antikink at the origin and
annihilate, producing a kink with an excited discrete mode.  The
system oscillates back to $K\bar{K}K$, but the separation is reduced.
The motion continues, essentially as a large amplitude
nonlinear oscillation of the discrete mode.  Energy is slowly
transferred to radiation modes, which is emitted symmetrically and
escapes to infinity.

\begin{figure}[htc]
\centerline{ \epsfxsize=10cm \epsfbox{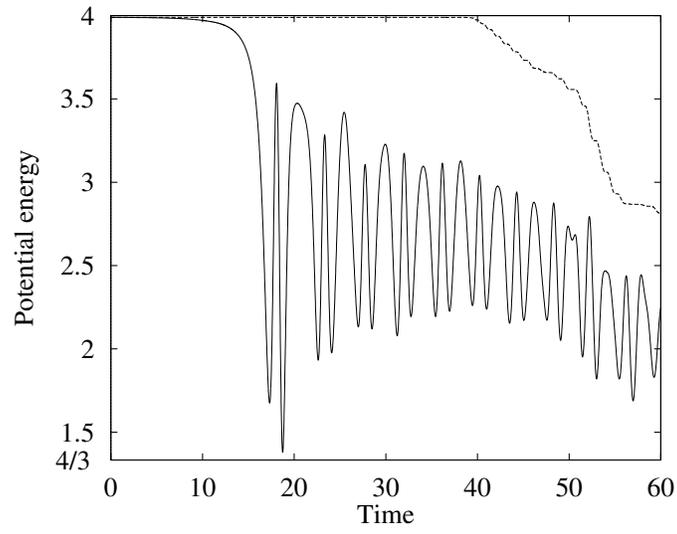} }
\caption{ Time dependence of the potential energy (solid)
and the total energy (dashed). }
\label{potfull:fig}
\end{figure}

Fig. \ref{potfull:fig} shows the potential energy and total energy as a
function of time.  Total energy is conserved until $t\simeq 40$, which
is when radiation first arrives at the boundary of the simulation interval.
Our boundary conditions absorb this radiation. Notice that at $t=
18.75$, after one oscillation of the
field, the potential energy is about 1.38 which is very close
to ${4 \over 3}$, the energy of the static kink, so the field dynamics
passes very close to the bottom of the valley at that time. Notice
also the very long
time during which the system oscillates with frequency just less than
$\sqrt{3}$, and the slow production of radiation.  In one period
of oscillation the potential energy has two minima but these have unequal
spacing in time. We have estimated
theoretically the rate of energy loss to radiation from the
oscillating discrete mode, and this agrees well with what we see
numerically (see Appendix A).

Fig. \ref{ampl:fig} shows $A(t)$, the amplitude of the discrete mode
calculated using eq.(\ref{amplitude}), for our dynamical field $\phi (x,t)$.
If the $K\bar{K}K$  system moved exactly in the valley, with total
energy less than 4, $A$ would oscillate
without loss of amplitude.  We see in Fig. \ref{ampl:fig}
that $A$ oscillates with slowly decreasing amplitude.
The asymmetry of the oscillation can be
understood from the asymmetry of the potential about the bottom of
the valley.  Note that the first time $A$ is positive, it
reaches 1.15.  This is close to the value 1.3 predicted
for motion in the valley with total energy 4.  Fig. \ref{toppot:fig}
shows the field
$\phi$ obtained dynamically, at the moment when $A$ as defined by
(\ref{amplitude}) is 1.15, compared to the ansatz (\ref{breather})
with $A = 1.15$.

Second, we have considered the initial data $\phi = \phi_K (x),
\dot{\phi} = -C \eta_D (x)$, that is, a single kink with its discrete
mode excited.  This motion
starts at the bottom of the valley and tangent to the valley bottom, and
although the valley is curved, we expect the motion to be guided
round the curve, so that a $K\bar{K}K$ configuration
will be produced if $C$ is sufficiently large.

If $C = 2 \sqrt{2}$, then the initial kinetic energy is ${8 \over 3}$, which
is just sufficient to produce a kink-antikink pair.
In practice, some energy goes into radiation, and this value of
$C$ is not large enough.  The critical value for kink-antikink
pair production is $C\simeq 3.06$.  Here, two kinks reach
the boundary of our region $-20 \leq x \leq 20$, leaving an antikink
at the origin, and since the forces
between the kinks and the antikink are negligible at these
separations, the kinks would presumably travel to infinity.  Since 3.06 is not
much greater than $2\sqrt{2}$ , we see that exciting the discrete mode of a
single kink is an efficient mechanism for producing a $K\bar{K}$ pair.
As $C$ increases beyond 3.06, so the outgoing kinks have higher
speed, and there is also more radiation.
Energetically, two $K\bar{K}$ pairs could be produced when $C=4$ but
in practice, more energy is needed, and the production first occurs
when $C=4.71$.

Figs. \ref{3kinks-a:fig} and  \ref{3kinks-b:fig} show the field
$\phi (x,t)$ for $-20 \leq x \leq 20$ and $0\leq t \leq 60$ with $C = 3.06$.
One can see how the discrete mode evolves into the $K\bar{K}K$ system.
The small amount of associated radiation is visible too.
Fig. \ref{5kinks-b:fig}  shows the creation of two $K\bar{K}$ pairs
when $C=4.71$.

\section{Conclusion}

The $K\bar{K}K$ system with reflection symmetry
is well-described by a one-dimensional reduced dynamical system with
an asymmetric potential. This is quite surprising because annihilation
of a kink and antikink
can and does occur, releasing a large amount of energy.  However, most
of this energy is converted to the oscillation of the discrete mode of
the remaining kink.  The collective coordinate of
the one-dimensional system can therefore be interpreted as the
separation of the kinks from the antikink, or as the amplitude of
the discrete mode, in different regions. We have defined the reduced
system on one side of the minimum of the potential by gradient flow,
on the other by an ansatz just involving the discrete mode,
but perhaps a better ansatz is possible. The reduced dynamics is
a good approximation to the full field dynamics, because the discrete
mode couples quite weakly to radiation modes.

We conjecture that in other field theories, multi-soliton and
soliton-antisoliton
dynamics can be similarly reduced to finite-dimensional systems using
gradient flow, and
that discrete vibrational modes of the solitons may again play an important
role. Soliton-antisoliton pair production may be associated with large
amplitude excitations of the discrete modes in these other field theories.
Perhaps soliton-antisoliton pair production at high temperatures can
be catalysed by the presence of solitons, if these solitons have
suitable discrete vibrational modes.

%


\newpage

\appendix
\section*{Appendix A}
\section*{\it Radiation from an oscillating kink}
\setcounter{section}{1}
\setcounter{equation}{0}

In the linear approximation, the discrete mode of the kink oscillates
harmonically with frequency $\sqrt{3}$. However, when the
nonlinearity of the fields is taken into account, this mode couples
to the continuum radiation modes and eventually all its energy
is radiated away. We can estimate the rate at which this happens
as follows.

Let us write the field $\phi$ as
\begin{equation}\label{kink-field:Ap}
\phi (x,t) = \phi_K (x) + A(t) \ \eta_D (x) + \eta (x,t)
\end{equation}
where $\phi_K (x) = \tanh (x)$ and $\eta_D (x) = \sinh (x)/ \cosh^2(x)$.
The function $\eta (x,t)$ represents the continuum part of the deformation
of the kink, and hence is orthogonal to the discrete mode
\begin{equation}\label{orthogonality:Ap}
\int^{\infty}_{-\infty} \eta (x,t) \eta_D (x) \mbox{d} x = 0.
\end{equation}
It is consistent to assume $\eta(x,t)$ is odd in $x$. If we substitute
(\ref{kink-field:Ap}) into the field equation (\ref{full-eq})
we find
\begin{eqnarray}\label{field-eq:Ap}
(\ddot{A} + 3 A) \eta_D +
\ddot{\eta}  - \eta'' + (6 \phi_K^2 - 2) \eta &=&
- 6 (\eta + \phi_K) \eta_D^2 A^2 \nonumber \\
& & - 6 (\eta + 2 \phi_K) \eta \eta_D A  \nonumber \\
& & - 2 \eta_D^3 A^3 - 6 \phi_K \eta^2 - 2 \eta^3
\end{eqnarray}
At ${\cal O}(A)$, the discrete mode oscillates with frequency
$\sqrt{3}$, and there is no source for $\eta$, so it is consistent
to set $\eta=0$. At  ${\cal O}(A^2)$, there is a source for
$\eta$, the term $- 6 \phi_K \eta_D^2 A^2$. If $A$ is small,
we may suppose that $\eta$ is  ${\cal O}(A^2)$, and
we may neglect terms in (\ref{field-eq:Ap}) involving $A^3$, $\eta^2$,
$\eta A$, etc. The reduced system is
\begin{equation}\label{O(A**2):Ap}
(\ddot{A} + 3 A) \eta_D +
\ddot{\eta}  - \eta'' + (6 \phi_K^2 - 2) \eta =
- 6 \phi_K \eta_D^2 A^2
\end{equation}
The source term $- 6 \phi_K \eta_D^2 A^2$ splits into two parts,
a projection onto the discrete mode, and a projection orthogonal to
this. The projection onto the discrete mode implies an anharmonic
oscillation of $A$, but this is not a large effect and the frequency
of the oscillation is unchanged to lowest order. The projection
orthogonal to the discrete mode is the source for $\eta$.
Fortunately the eigenfunctions of the Schr\"{o}dinger operator
\begin{equation}\label{Sch-Op:Ap}
- {\mbox{d}^2 \over \mbox{d}x^2} + (6 \phi_K^2 - 2)
\end{equation}
are known exactly \cite{Rajaraman82}, so the Green's function can be computed
and the response of $\eta$ to the source can be calculated by explicit
integration.

The truncation (\ref{O(A**2):Ap}) is not energy conserving.
The discrete mode oscillation creates radiation but there is no
backreaction on the amplitude of the discrete mode. We may calculate
the backreaction by finding the energy carried away by
the radiation, and may infer the rate of decrease of the amplitude
of the discrete mode, using energy conservation. Our calculation
depends on the rate of energy loss being small during
the period ${2\pi \over \sqrt{3}}$.

Let the projection of $\phi_K \eta_D^2$ onto the discrete mode
be $\alpha \eta_D$, so the orthogonal projection is
$\phi_K \eta_D^2 - \alpha \eta_D$. The constant $\alpha$
is determined by
\begin{equation}\label{alpha:Ap}
         \int^{\infty}_{-\infty} \phi_K (x) \eta_D (x)^3 \mbox{d}x
- \alpha \int^{\infty}_{-\infty} \eta_D (x)^2 \mbox{d}x = 0
\hspace{0.3cm} ,
\end{equation}
implying $\alpha = {3\pi \over 32}$. The oscillation of the discrete
mode, to ${\cal O}(A^2)$, therefore obeys
\begin{equation}\label{A-O(A**2):Ap}
\ddot{A} + 3 A + 6\alpha A^2 = 0
\end{equation}
which agrees with what would be obtained, at this order,
from (\ref{A-Kin}) and (\ref{A-Pot}).
The equation for $\eta$ is
\begin{equation}\label{eta-O(A**2)-1:Ap}
\ddot{\eta}  - \eta'' + (6 \phi_K^2 - 2) \eta =
6\ (\alpha \eta_D  - \phi_K \eta_D^2)\ A^2
\end{equation}
Let us now assume that $A = A_0 \cos (\sqrt{3} t)$, so
$A^2 = {1 \over 2} A_0^2 (\cos (2 \sqrt{3} t) + 1)$.
The response of $\eta$ to the time-independent source is
itself time-independent and carries away no energy. The important
part of $\eta$ is at the single frequency $2 \sqrt{3}$. Let us therefore
consider the equation
\begin{equation}\label{eta-O(A**2)-2:Ap}
\ddot{\eta}  - \eta'' + (6 \phi_K^2 - 2) \eta =
f (x) \exp (i \omega t)
\end{equation}
where $\omega > 2$. Setting $\eta (x,t) = \eta (x) \exp (i \omega t)$,
we obtain
\begin{equation}\label{eta-O(A**2)-3:Ap}
- \eta'' + (6 \phi_K^2 - 2 - \omega^2) \eta = f (x) \ .
\end{equation}
The homogenous equation (\ref{eta-O(A**2)-3:Ap}) (with $f \equiv 0$)
has exact solutions \cite{Rajaraman82}
\begin{eqnarray}\label{eta-homo:Ap}
\eta_q (x) &=& (3 \phi_K^2(x) - 1 - q^2 - 3 i q \phi_K (x) ) \exp (+ i q x)
\nonumber \\
\eta_{-q} (x) &=& (3 \phi_K^2(x) - 1 - q^2 + 3 i q \phi_K (x)) \exp (- i q x)
\end{eqnarray}
where $q =\sqrt{\omega^2 - 4}$.
 From the form of these solutions as $x \to \infty$,
we compute the Wronskian
\begin{equation}\label{wronskian:Ap}
W = \eta_q (x)\ \eta'_{-q} (x) - \eta'_q (x)\ \eta_{-q} (x) =
- 2 i q (q^2 + 1) (q^2 + 4) \ .
\end{equation}
The solution of (\ref{eta-O(A**2)-3:Ap}) that we want is the one with
outgoing radiation. The relevant Green's function is
\begin{equation}\label{green:Ap}
G (x,\xi) =
\left\{ \begin{array}{ll}
- \displaystyle {1 \over W} \eta_{-q} (\xi)\ \eta_q (x) & \ \ (x < \xi) \\
- \displaystyle {1 \over W} \eta_{q} (\xi)\ \eta_{-q} (x) & \ \ (x >
\xi) \ .
\end{array}
\right.
\end{equation}
So the desired solution of (\ref{eta-O(A**2)-3:Ap}) is
\begin{equation}\label{eta-sol:Ap}
\eta (x)  =   - {1 \over W} \eta_{-q} (x)
\int_{-\infty}^{x} f (\xi)  \eta_{q} (\xi) \mbox{d}\xi \
- {1 \over W} \eta_{q} (x)
\int_{x}^{\infty} f (\xi)  \eta_{-q} (\xi) \mbox{d}\xi \ .
\end{equation}
We may use the asymptotic form of (\ref{eta-homo:Ap})
to obtain the form of the solution (\ref{eta-sol:Ap}) for
large $x$,
\begin{equation}\label{eta-asy-1:Ap}
\eta (x)  =   \frac{1}{2 i q (2 - q^2 - 3 i q)}
\exp (- i q x)
\int_{-\infty}^{\infty} f (\xi)  \eta_{q} (\xi) \mbox{d}\xi \ .
\end{equation}
Thus, for the source  $-3 A_0^2 \phi_K \eta_D^2 \cos (2 \sqrt{3} t)$,
$\eta(x,t)$ is asymptotically
\begin{equation}\label{eta-asy-2:Ap}
\eta (x,t)  =  \mbox{Re} \left ( \frac{- 3 A_0^2}{2 i q (2 - q^2 - 3 i q)}
\exp i (\omega t - q x)
\int_{-\infty}^{\infty} \phi_K (\xi) \eta_D^2 (\xi)  \eta_{q} (\xi)
\mbox{d}\xi
\right )
\end{equation}
where $\omega = 2 \sqrt{3}$ and $q = 2 \sqrt{2}$. It is sufficient
here to write the source as proportional to $\phi_K \eta_D^2$,
since the integral of $\eta_D \eta_q$ vanishes.

Eq.(\ref{eta-asy-2:Ap}) represents outward moving radiation for large $x$.
There is similar outward radiation for large negative $x$.
To calculate the integral in (\ref{eta-asy-2:Ap}), we use the result
\begin{eqnarray}\label{integral:Ap}
I (q) & = &  \int_{-\infty}^{\infty} \phi_K (\xi) \eta_D^2 (\xi)
(3 \phi_K^2 (\xi) - 1 - q^2 - 3 i q \phi_K (\xi)) \exp (i q \xi) \mbox{d}\xi
\nonumber \\
& = &  \frac{i \pi q^2 }{48 \sinh(\pi q/2)} (q^2 + 4) (q^2 - 2) \ .
\end{eqnarray}
Thus (\ref{eta-asy-2:Ap}) reads now
\begin{equation}\label{eta-asy-3:Ap}
\eta (x,t)  =
 \frac{\pi q (q^2 - 2)}{32 \sinh (\pi q/2)}
\sqrt{\frac{q^2 + 4}{q^2 + 1}}\ A_0^2 \cos (\omega t - q x - \delta)
\end{equation}
where $\delta$ is a phase depending on $q$. We are
only interested in the amplitude of the radiation, not its phase.
For $q = 2 \sqrt{2}$
the amplitude is
\begin{equation}\label{eta-amp:Ap}
R  =  \frac{\pi \sqrt{3/8}}{\sinh (\pi \sqrt{2})} A_0^2 = 0.0453\ A_0^2
\end{equation}
For a wave of the form $ \eta = R \cos (\omega t - q x - \delta)$ the
average energy flux to the right is ${1\over 2} R^2 \omega q$, so
the total energy flux away from the oscillating kink is $R^2 \omega q$.
The rate of energy loss by the discrete mode (averaged over a period) is
therefore
\begin{equation}\label{rate-E:Ap}
\frac{\mbox{d} E}{\mbox{d} t}  =
- (0.0453)^2 4 \sqrt{6}\ A_0^4 = - 0.020\ A_0^4 \ .
\end{equation}
On the other hand, the energy of the discrete mode is $E = A_0^2$, so
the rate of decay of the amplitude is given by
\begin{equation}\label{rate-A:Ap}
\frac{\mbox{d} A_0^2}{\mbox{d} t}  = - 0.020\ A_0^4
\end{equation}
which can be integrated to give
\begin{equation}\label{A-decay:Ap}
\frac{1}{A_0^2(t)} - \frac{1}{A_0^2(0)} = 0.020\ (t - t_0) \ .
\end{equation}

We can calculate the decay of the discrete mode amplitude due to
radiation in another way. Consider again the exact equation
(\ref{field-eq:Ap}).

Using the orthogonality condition (\ref{orthogonality:Ap}),
we project the equation (\ref{field-eq:Ap}) onto the discrete mode
\begin{equation}\label{A:Ap}
\ddot{A} + \omega_0^2 A = a_0 + a_1 A + a_2 A^2 + a_3 A^3
\end{equation}
where we have defined
\begin{equation}\label{a-coef:Ap}
\begin{array}{ll}
a_0 = - 3 \displaystyle \int^{\infty}_{-\infty}
(3 \phi_K + \eta) \eta^2 \eta_D \mbox{d} x; &
a_1 = - 9 \displaystyle \int^{\infty}_{-\infty}
(2 \phi_K + \eta) \eta \eta_D^2 \mbox{d} x   \\
a_2 = - 9 \displaystyle \int^{\infty}_{-\infty}
\eta \eta_D^3 \mbox{d} x - 6 \alpha;        &
a_3 = - 3 \displaystyle \int^{\infty}_{-\infty}
\eta_D^4 \mbox{d} x = - {12\over 35} \ .
\end{array}
\end{equation}
Following ref.\cite{Lan/LifT1}, and as in \cite{Segur83},
we seek a perturbative solution
for $\phi(x,t)$ of the form (\ref{kink-field:Ap}) with
\begin{eqnarray}\label{perturbation:Ap}
A &=&  A^{(1)} + A^{(2)} + A^{(3)} + \cdots \nonumber \\
\eta &=& \eta^{(1)} + \eta^{(2)} + \eta^{(3)} + \cdots
\end{eqnarray}
and with $A$ having period $2\pi/\tilde{\omega}$, where
\begin{eqnarray}\label{omega:Ap}
\tilde{\omega} &=& \omega_0 + \omega^{(1)} + \omega^{(2)} + \cdots
\end{eqnarray}
At the first order, the discrete mode oscillates with the frequency
$\omega_0=\sqrt{3}$, and there is no source term for the continuum
modes, so $A^{(1)}=A_0 \cos(\omega_0 t)$  and $\eta^{(1)}=0$ .

At the second order, eq.(\ref{A:Ap}) reduces to
\begin{equation}\label{A-2:Ap}
\ddot{A}^{(2)} + \omega_0^2 A^{(2)} = -6 \alpha {A^{(1)}}^2 +
2 \omega_0 \omega^{(1)} A_0 \cos(\omega_0 t) \ .
\end{equation}
The requirement that the resonance term (at frequency $\omega_0$)
be absent implies $\omega^{(1)}=0$, so we obtain exactly
eq.(\ref{A-O(A**2):Ap}).
The latter displays the first anharmonic correction to the
discrete mode oscillation.
Writing ${A^{(1)}}^2 = {1\over 2} A_0^2 (\cos(2 \omega_0 t) + 1)$,
the solution of eq.(\ref{A-2:Ap}) reads
\begin{equation}\label{A-2-sol:Ap}
A^{(2)} = {\alpha \over {\omega_0}^2} A_0^2
(\cos(2 \omega_0 t) - 3) \ .
\end{equation}
The correction (\ref{A-2-sol:Ap}) tells us that the main source
of the asymmetric oscillation of $A$ in Fig. \ref{ampl:fig}
is the cubic term of the  potential (\ref{A-Pot}).

By expanding eq.(\ref{field-eq:Ap}) up to the second order
in $\eta$ and $A$, we obtain
\begin{equation}\label{eta-2:Ap}
{\ddot{\eta}}^{(2)} - {\eta''}^{(2)} +
(6 \phi_K^2 - 2) \eta^{(2)} =
 6 (\alpha - \phi_K \eta_D) \eta_D {A^{(1)}}^2
\end{equation}
which is exactly eq.(\ref{eta-O(A**2)-1:Ap}), therefore $\eta^{(2)}$
is given asymptotically by (\ref{eta-asy-3:Ap}).

>From eq.(\ref{A-2:Ap}) we see no $\eta$ contribution to the $A^{(2)}$
correction, consequently there is no energy conservation at this order.
To find the modification to the discrete mode oscillation caused
by radiation, we must go to the next order. The third order part of
(\ref{A:Ap}) is given by
\begin{eqnarray}\label{A-3:Ap}
\ddot{A}^{(3)} + \omega_0^2 A^{(3)} &=&
(6 i q \tilde{R}^2 -  6 {\alpha^2\over \omega_0^2} + {a_3\over 4})
A_0^3 \cos (3 \omega_0 t)  \nonumber \\ & &
+ (2 \omega_0 \omega^{(2)} + 6 i q \tilde{R}^2 A_0^2 +
30 {\alpha^2\over \omega_0^2} A_0^2 +
{3\over 4} a_3 A_0^2) A_0 \cos (\omega_0 t)
\end{eqnarray}
where $q = 2\sqrt{2}$ and  $\tilde{R}=R/A_0^2$ with $R$ given by
(\ref{eta-amp:Ap}). Note the appearance of complex numbers
resulting from the coupling of the discrete mode with the imaginary
part of the continuum (the real part doesn't contribute at this order).
The condition for the resonance term to be absent is
\begin{eqnarray}\label{omega-2:Ap}
\omega^{(2)} &=& - (30 {\alpha^2\over \omega_0^2} +
{3\over 4} a_3 + 6 i q \tilde{R}^2) \frac{A_0^2}{2 \omega_0} \nonumber
\\ &=& - (0.176 + i 0.010) A_0^2
\end{eqnarray}
The second order real correction to the frequency is negative,
so the period of the discrete mode oscillation is greater than
$2\pi/\omega_0$. This effect is confirmed by Fig. \ref{ampl:fig}
where we see a longer period at earlier times, when the discrete
mode amplitude is large.
The imaginary correction to the frequency implies a decay of the
amplitude, the effect we are seeking, which in turn means that the real
frequency increases with time.
The decay rate of the amplitude is given by
\begin{equation}\label{A-3-sol:Ap}
\frac{\mbox{d} A_0}{\mbox{d} t}  = - 0.010\ A_0^3
\end{equation}
which is equivalent to (\ref{rate-A:Ap}).
Fig. \ref{amplpert:fig} shows the evolution of $\phi (x,t)$ calculated
numerically with initial data
$\phi(x,t)  = \phi_K (x) + \eta_D (x)$, so $A_0(0) = 1$.
The dashed line shows the amplitude calculated up to the second-order,
and  including the frequency correction (\ref{omega-2:Ap}).
The decay of the amplitude of the discrete mode is thus seen to be
well approximated by (\ref{A-decay:Ap}).

\newpage


\newpage

\section*{Figure captions}

\vspace{1cm}
\noindent Figure 1:  The gradient flow annihilation process for the
$K\bar{K}K$ system at time 0 (solid), 59 (dotted), 59.64 (dashed)
and 61 (dashed dotted).

\vspace{1cm}
\noindent Figure 2: Time dependence of the potential energy along
the gradient flow curve.

\vspace{1cm}
\noindent Figure 3: Arc length dependence of the potential energy
in the valley. Gradient flow (solid) and discrete mode results for $s<0$
(the discrete mode result for $s>0$ is also shown).

\vspace{1cm}
\noindent Figure 4: The full dynamics annihilation process.

\vspace{1cm}
\noindent Figure 5: Time dependence of the potential energy (solid)
and the total energy (dashed).

\vspace{1cm}
\noindent Figure 6: Time dependence of the discrete mode amplitude.

\vspace{1cm}
\noindent Figure 7: The field $\phi$ (solid) compared to the discrete mode
ansatz at $A =1.15$ (dashed).

\vspace{1cm}
\noindent Figure 8: Creation process of three kinks for $C=3.06$.

\vspace{1cm}
\noindent Figure 9: Snapshots of the three kinks creation process
at time 10 (solid), 30 (dotted), 60 (dashed) and 90 (dashed dotted).

\vspace{1cm}
\noindent Figure 10: Snapshots of the five kinks creation process
at time 10 (solid), 20 (dotted), 40 (dashed) and 60 (dashed dotted).

\vspace{1cm}
\noindent Figure 11: Time dependence of the discrete mode amplitude
 with $A_0(0) = 1$ (solid) compared to the second order approximation
(dashed).

\begin{figure}[htc]
\centerline{ \epsfxsize=10cm \epsfbox{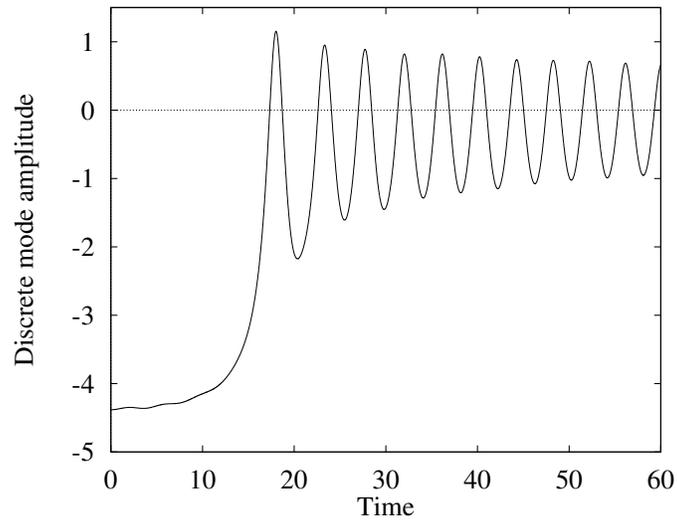} }
\caption{ Time dependence of the discrete mode amplitude. }
\label{ampl:fig}
\end{figure}

\begin{figure}[htc]
\centerline{ \epsfxsize=10cm \epsfbox{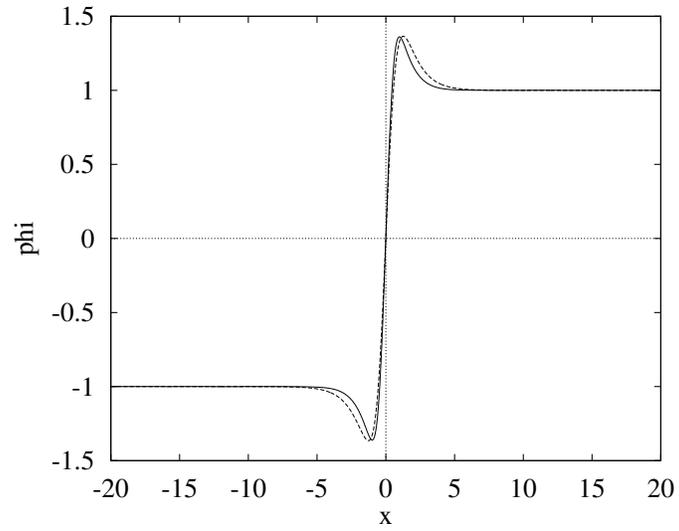} }
\caption{ The field $\phi$ (solid) compared to the discrete mode
ansatz at $A =1.15$ (dashed). }
\label{toppot:fig}
\end{figure}

\begin{figure}[htc]
\setlength{\unitlength}{1in}
\begin{picture}(4.6,2.3)(0,0)
\put(0.9,-0.2){\epsfbox{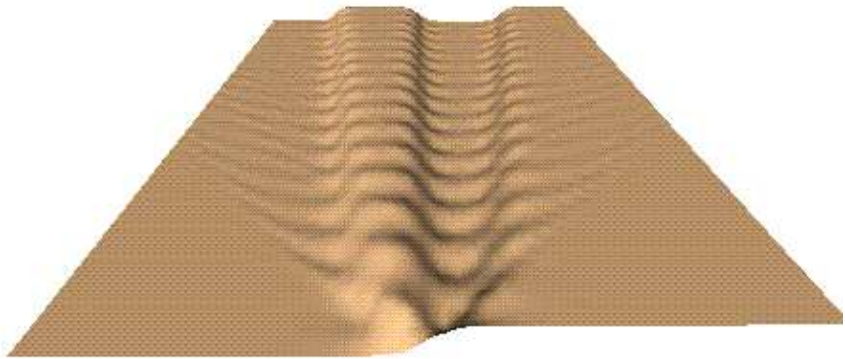}}
\end{picture}
\caption{Creation process of three kinks for $C=3.06$.}
\label{3kinks-a:fig}
\end{figure}

\begin{figure}[htc]
\centerline{ \epsfxsize=10cm \epsfbox{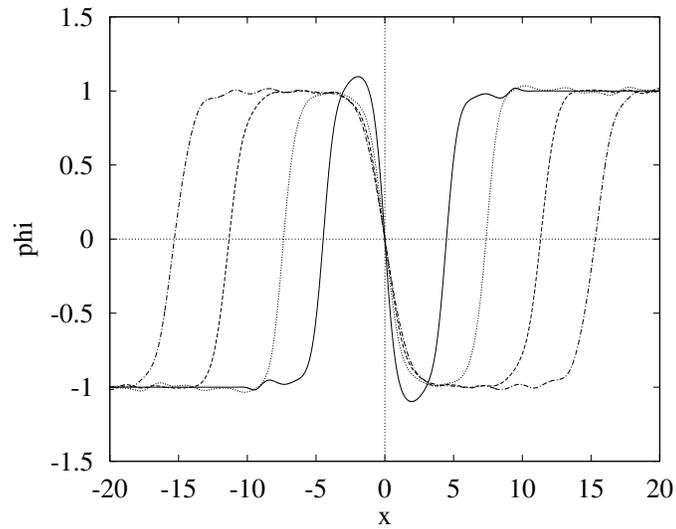} }
\caption{Snapshots of the three kinks creation process at time 10
(solid), 30 (dotted), 60 (dashed) and 90 (dashed dotted).}
\label{3kinks-b:fig}
\end{figure}

\begin{figure}[htc]
\centerline{ \epsfxsize=10cm \epsfbox{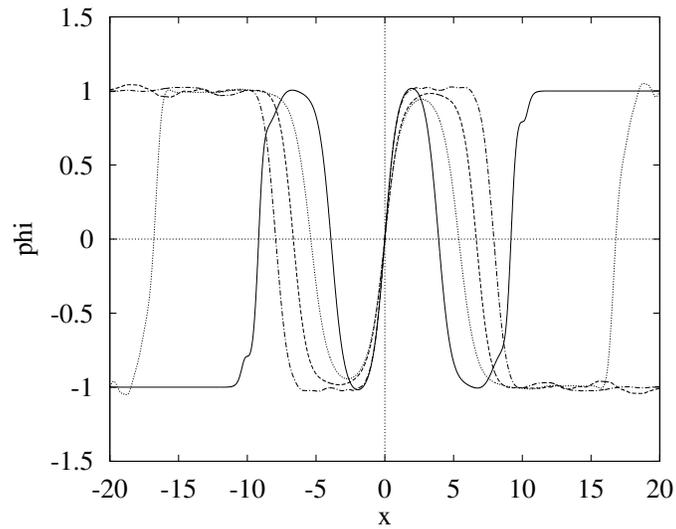} }
\caption{Snapshots of the five kinks creation process at time 10
(solid), 20 (dotted), 40 (dashed) and 60 (dashed dotted).}
\label{5kinks-b:fig}
\end{figure}

\begin{figure}[htc]
\centerline{ \epsfxsize=10cm \epsfbox{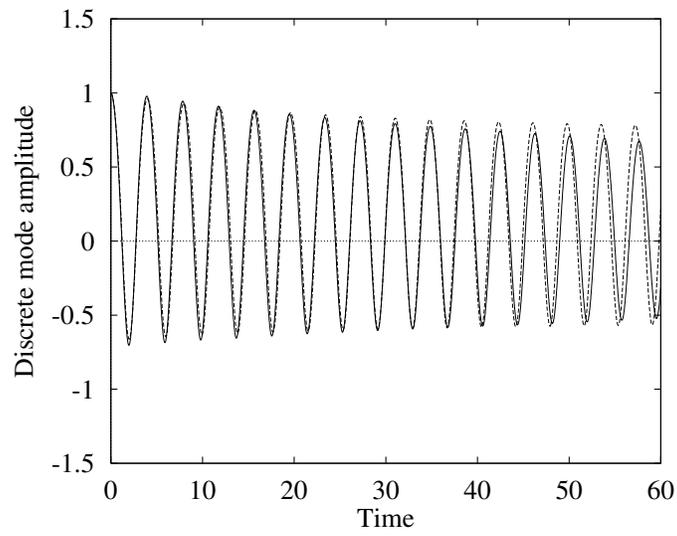} }
\caption{Time dependence of the discrete mode amplitude with $A_0(0) = 1$
(solid) compared to the second order approximation (dashed). }
\label{amplpert:fig}
\end{figure}

\end{document}